\documentclass[twocolumn,prd,nofootinbib,superscriptaddress,eqsecnum,tightenlines,8pt]{revtex4}

\usepackage{fancyhdr}
\usepackage{amsmath,amsfonts,amssymb}
\usepackage[colorlinks,citecolor=blue,linkcolor=blue,urlcolor=blue]{hyperref}
\usepackage[english]{babel}

\usepackage{color}
\definecolor{red}{rgb}{1,0,0}
\definecolor{gre}{rgb}{0,0.6,0}
\definecolor{blu}{rgb}{0,0,1}

\usepackage{mathenv}
\usepackage{ulem}

\usepackage[pdftex]{graphicx}
\usepackage{mathrsfs} 
\DeclareGraphicsExtensions{.jpg, .JPG , .png , .pdf, .gif}
\def\be{\begin{equation}}
\def\ee{\end{equation}}

\begin{document}

\title{A first step towards the inflationary trans-planckian problem treatment\\ in Loop Quantum Cosmology}

\date{\today}

\author{Killian Martineau}
\email{martineau@lpsc.in2p3.fr}
\affiliation{
Laboratoire de Physique Subatomique et de Cosmologie, Universit\'e Grenoble-Alpes, CNRS-IN2P3\\
53, Avenue des Martyrs, 38026 Grenoble cedex, France\\
}%

\author{Aur\'elien Barrau}
\email{Aurelien.Barrau@cern.ch}
\affiliation{
Laboratoire de Physique Subatomique et de Cosmologie, Universit\'e Grenoble-Alpes, CNRS-IN2P3\\
53, Avenue des Martyrs, 38026 Grenoble cedex, France\\
}%

\author{Julien Grain}
\email{julien.grain@ias.u-psud.fr }
\affiliation{Institut d'astrophysique spatiale, Universit\'e Paris-Sud, CNRS \\
B\^atiments 120 \`a 121, Universit\'e Paris Sud, 91405 ORSAY, France
}%

\begin{abstract} 
For most initial conditions, cosmologically relevant physical modes were trans-planckian at the bounce time, often by many magnitude orders. We improve the usual loop quantum cosmology calculation of the primordial power spectra -- in the inflationary framework -- by accounting for those trans-planckian effects through modified dispersion relations. This can induce drastic changes in the spectrum, making it either compatible or incompatible with observational data, depending on the details of the choices operated.
\end{abstract}

\maketitle

\section{Introduction}

Even though loop quantum gravity (LQG) is one of the most studied attempt to non-perturbatively and consistently quantize general relativity (GR), the derivation of fully clear experimental predictions is still missing. Together with black holes (see, {\it e.g.}, \cite{Rovelli:2014cta,Barrau:2014hda,Haggard:2014rza,Barrau:2014yka,Barrau:2015uca,Barrau:2015ana,Barrau:2016fcg,Barrau:2016qri} for recent studies), cosmology is the natural sector of the theory to consider to search for links with observations (see, {\it e.g.}, \cite{Wilson-Ewing:2016yan} and references therein for a recent and comprehensive review). Many important results have been obtained in loop quantum cosmology (LQC). In particular, primordial spectra have been calculated in the so-called dressed metric approach (see, {\it e.g.}, \cite{Agullo3}), in the deformed algebra approach (see, {\it e.g.}, \cite{eucl3} and, {\it e.g.} \cite{Bolliet:2015bka} for a comparison of both approaches), and in the separate universe approach (see, {\it e.g.}, \cite{Wilson-Ewing:2015sfx}).\\

It is fair to say that the background dynamics is now well established in LQC (see, {\it e.g.}, \cite{lqc9} and references therein). The well known modified Friedmann equation is not only confirmed by numerical simulations  \cite{Diener:2013uka}, but also (mostly) by new calculations in group field theory (GFT) \cite{Oriti:2016qtz,} (building on \cite{Gielen:2013kla}), and in quantum reduced loop gravity (QRLG)  \cite{Alesci:2014rra}.\\

The situation is however quite different for perturbations. Not only because, as previously mentioned, different settings are being considered but also -- and most importantly -- because trans-planckian effects are mostly neglected or, to say the least, not treated as a dominant process. It is our feeling that too much emphasis has been put on {\it density} effects and not enough on {\it length} effects. The former do indeed trigger the bounce but the latter cannot be ignored in a theory that basically states that there is nothing smaller than the Planck length\footnote{This is actually more complicated than this as the length operator is not as well defined as the area or volume operator in LQG. There are several consistant proposals \cite{Thiemann:1996hw,Bianchi:2008es,Ma:2010fy} but a totally unambiguous picture is still missing.}. As soon as the number of e-folds of inflation is higher than 70, all modes of physical interest were trans-planckien ({\it i.e.} their physical length was smaller than the Planck length) at the bounce time. This is a well known cosmological problem \cite{Martin:2000xs} but the specific setting of LQG imposes to take a special look at this question. This is especially true as the number of e-folds can, to some extent, be ``predicted" in LQC when the Universe is filled with a scalar field yielding a phase in inflation. There is no consensus on this value (see \cite{AS2011} for a school of thought, \cite{bl} for another one, and \cite{Martineau:2017sti} for a general study) but it is anyway generically much higher that 70. If one takes a quite ``standard" GR value like $N=10^{13}$, it means that modes seen in the CMB were approximately $e^{10^{13}}$ times smaller than the Planck length at the bounce time ! Even with smaller values, like $N=145$ as favored by the specific dynamics of LQC \cite{bl}, the relevant lengths are typically $10^{33}$ times smaller than the Planck length at the bounce time. This cannot be ignored. This problem is obviously related with inflation -- and can be evaded in this framework only by assuming arbitrarily the smallest possible value $N\sim70$ -- but does not appear in the matter bounce scenario \cite{WilsonEwing:2012pu} or in the ekpyrotic setting \cite{Cailleteau:2009fv}.\\

As a first elementary step to account for trans-planckian effets we suggest to use modified dispersion relations (MDRs) in the LQC framework. This is obviously not the final word on this question. For sure, trans-planckian effects should ideally be considered in the full theory (following for example \cite{Ashtekar:2009mb}) but we believe that MDRs are a meaningful first step. This is at least a way to take into account the trans-planckian problem, which is especially interesting in a bouncing-universe framework as MDRs are known to have, in this case, possible observational consequences without involving negative energies \cite{Brandenberger:2002ty,Brandenberger:2012aj} or arbitrary vacuum states.\\

In this article, we study the effects of two different well motivated MDRs on the primordial spectra obtained both in the deformed algebra and in the dressed metric approach to perturbations in LQC. The MDRs we consider in this work both have an energy-momentum dependance which grows less (or even decreases) than the usual linear behavior in the Planckian regime. Intuitively, from the statistical viewpoint, this means that the energy required to excite trans-planckian (from the momentum viewpoint) modes is less than naively expected. From the microscopic viewpoint this is just a very rudimentary way of taking into account the granular structure of spacetime. We do not consider in this study Corley-Jacobson MDRs \cite{Corley:1996ar} as the frequency either becomes negative for $k>k_{Pl}$ (case $\omega^2 = k^2-k^4 / k_0^4$) -- raising serious stability issues and being hard to trust in the trans-plackian regime which is critical for this study -- or rises faster with $k$ (cubic model), making the situation worst than usual.

\section{Background dynamics in Loop Quantum Cosmology}

\subsection{The two main equations}

The background evolution equation in LQC is given by 

\begin{equation}
H^{2} = \dfrac{\kappa}{3} \rho \left( 1 - \dfrac{\rho}{\rho_{c}} \right),
\label{FriedmannLQC}
\end{equation}
where $\kappa = 8 \pi G$, $H = \dot{a}/a$ stands for the Hubble parameter, $\rho$ is the energy density, and $\rho_{c} = 0,41 \rho_{\text{Pl}}$ is a maximum energy density. This equation obviously leads to a bounce. We assume  the Universe to be filled with a massive scalar field $\Phi$, that is a potential $V(\Phi) = m^{2} \Phi^{2}/2$.  This potential is not favoured by Planck data but is useful for comparison with other studies and the main argument of this study is not potential shape dependant. 

The evolution of the field is given by the Klein-Gordon equation:

\begin{equation}
\ddot{\Phi} + 3 H \dot{\Phi} + \dfrac{dV(\Phi)}{d\Phi} =0.
\label{Klein-Gordon}
\end{equation}
With this content, and using the continuity equation
\begin{equation}
\dot{\rho} = - 3 H (\rho + P),
\end{equation}
$P$ being the pressure in the rest-frame of the fluid, the derivative of Eq. (\ref{FriedmannLQC}) becomes

\begin{equation}
\dot{H} = -\dfrac{\kappa}{2} \dot{\Phi}^{2} \left[1-2\dfrac{\rho}{\rho_{c}} \right]~.
\label{Raychauduuri}
\end{equation}
The Klein-Gordon equation (\ref{Klein-Gordon}), together with  Eq. (\ref{Raychauduuri}), constitute the set of equations that drives the background dynamics.  

\subsection{The background initial conditions}
  
The scalar field evolution can be described by two variables (related one to the other), the potential energy parameter $x$ and the kinetic energy parameter $y$, defined by

\begin{equation}
x(t) = \dfrac{m \Phi(t)}{\sqrt{2 \rho_{c}}} ~, ~~~~~ y(t) := \dfrac{\dot{\Phi}}{\sqrt{2\rho_{c}}}~,
\end{equation}

that satisfy

\begin{equation}
x^{2}(t)+y^{2}(t)=\dfrac{\rho(t)}{\rho_{c}} ~.
\end{equation}

The Klein-Gordon equation (\ref{Klein-Gordon}) can then be recast into a set of two first order ordinary differential equations (ODE):

\begin{equation*}
  \left\{
    \begin{aligned}
&  \dot{x}(t) = m y(t)~, \\
& \dot{y}(t) = -3H(t)y(t)-mx(t)~.
    \end{aligned}
  \right.
  \label{Klein Gordon 2 eqs}
\end{equation*}

The ratio of the two time scales $1/m$ and $1/\sqrt{3 \kappa \rho_{c}}$ is given by

\begin{equation}
\Gamma := \dfrac{m}{\sqrt{3 \kappa \rho_{c}}} ~.
\end{equation}

In the classical contracting phase, $\sqrt{\rho(t) / \rho_{c} } \ll \Gamma$ holds, thus the Klein-Gordon equation (\ref{Klein-Gordon}) reduces to a simple harmonic oscillator. In this case, $x$ and $y$ are well-approximated by

\begin{equation}
  \left\{
    \begin{aligned}
  x(t) \simeq \sqrt{\dfrac{\rho(0)}{\rho_{c}}} \sin(mt+\delta)~, \\
  y(t) \simeq \sqrt{\dfrac{\rho(0)}{\rho_{c}}} \cos(mt+\delta)~.
    \end{aligned}
  \right.
  \label{x(t) y(t) oscillants}
\end{equation}

The scalar field oscillation phase $\delta$ is not fundamentally relevant since the background dynamics is mostly independent of its value \cite{bl} (this statement is not true anymore when considering flat potentials \cite{Martineau:2017sti}). The energy density in the classical phase can be expressed as:

\begin{equation}
\rho(t) \simeq \rho_{c} \left(\dfrac{\Gamma}{\alpha}\right)^{2} \left[ 1 - \dfrac{1}{2 \alpha}  \left( mt + \dfrac{1}{2} \sin( 2 m t + 2 \delta) \right) \right]^{-2},
\label{rho(t)}
\end{equation}
where $\alpha=17/4 \pi +1$ is a free parameter set to ensure that $\sqrt{\rho(t) / \rho_{c} } \ll \Gamma$ remains valid and so that a sufficiently high number of field oscillations takes place in the contracting phase (more than 10), making it convenient to derive analytical solutions in the quadratic potential case.

To obtain the initial conditions for the matter sector one evaluates Eq. (\ref{x(t) y(t) oscillants}) and Eq. (\ref{rho(t)}) at $t=0$, leading to:

\begin{equation}
  \left\{
    \begin{aligned}
&  \Phi (0) = \sqrt{2 \rho(0)} \sin(\delta )/m~, \\
&  \dot{\Phi} (0) = \sqrt{2 \rho(0)} \cos(\delta )~,
    \end{aligned}
  \right.
\end{equation}
and
\begin{equation}
\rho(0) = \rho_{c} \left(\dfrac{\Gamma}{\alpha}\right)^{2} \left[ 1 - \dfrac{1}{4 \alpha} \sin(2 \delta) \right]^{-2} ~.
\end{equation}

For fixed values of $\alpha$, the only free variable which remains to be chosen to completely determine the initial parameters $\lbrace \Phi(0),\dot{\Phi}(0),\rho(0)\rbrace$ is the scalar field initial phase $\delta$. As previously mentioned, except if one performs an hyper-fine tuning, the dynamics is basically independent of its value, thus we arbitrarily choose $\delta=\pi/2$. \\

Nevertheless the choice of initial conditions for the background is not the key-point for this study and the main results do {\it not} depend on the particular initial conditions chosen.

\section{Perturbations on LQC background}

\subsection{The Mukhanov-Sasaki perturbations equation in the usual algebra}

The perturbed Einstein equations for a flat universe filled with a scalar field lead to the gauge-invariant Mukhanov-Sasaki equation:

\begin{equation}
v''(\eta, \vec{x}) - c_{s}^{2} \vartriangle v(\eta, \vec{x}) - \dfrac{z_{T/S}''(\eta)}{z_{T/S}(\eta)} v(\eta, \vec{x}) = 0 ~,
\label{Mukhanov}
\end{equation}
where $f^\prime$ denotes a derivative with respect to the conformal time $\eta$.
The canonical variable $v$ is a gauge-invariant combination of the metric coordinate perturbations (Bardeen variables) and of the scalar field perturbations. The variable $z$ depends on the kind of  perturbations considered and the indices $T/S$ refer to tensor or scalar modes.\\

The evolution of cosmological perturbations is equivalent to a free scalar field  $v$ with a time-dependent mass $m^{2} = - z_{T/S}''/z_{T/S}$ in a Minkowski space-time. The mass time-dependence is due to interactions between the perturbations and the expanding background (the energy is not conserved and they can be excited by borrowing energy from the expansion). 

The functions $v$ are promoted to be operators and the associated Fourier temporal mode functions satisfy

\begin{equation}
v_{k}''(\eta) + \left(k_{c}^{2} - \dfrac{z_{T/S}''(\eta)}{z_{T/S}(\eta)} \right) v_{k}(\eta) =0 ~,
\label{MukhanovSasaki Temporal modes}
\end{equation}

where $k_{c}$ refers to the considered comoving wavenumber.  As it will be important in the following to switch to physical wavenumbers, one rewrite this equation as

\begin{equation}
v_{k}''(\eta) + \left(a^{2}(\eta) k_{\varphi}^{2} - \dfrac{z_{T/S}''(\eta)}{z_{T/S}(\eta)} \right) v_{k}(\eta) =0,
\label{MukhanovSasaki Temporal modes physical k}
\end{equation}

where $k_{\varphi}(\eta) = k_{c}/a(\eta)$.\\

The aim of this study is to introduce modified dispersion relations in the Mukhanov-Sasaki equation, so as to account effectively for quantum length effects at the Planck scale. Those modified relations are relevant when applied to physical quantities and are therefore introduced in Eq. (\ref{MukhanovSasaki Temporal modes physical k}) by the replacement: $k_{\varphi}\rightarrow \mathcal{F}(k_{\varphi})$. The function $\mathcal{F}(k_{\varphi})$ depends on the chosen dispersion relation. The classical case corresponds to $\mathcal{F}(k_{\varphi})=k_{\varphi}$. The fundamental equation then becomes:

\begin{equation}
v_{k}''(\eta) + \left(a^{2}(\eta) \mathcal{F}(k_{\varphi})^{2} - \dfrac{z_{T/S}''(\eta)}{z_{T/S}(\eta)} \right) v_{k}(\eta) =0 ~.
\end{equation}

This can also be re-written in cosmic time:

\begin{equation}
    \begin{aligned}
& \ddot{v_{k}}(t) + H(t) \dot{v_{k}}(t)\\
&  + \left( \mathcal{F}(k_{\varphi})^{2} - \dfrac{\dot{z}_{T/S}(t)}{z_{T/S}(t)}H(t) - \dfrac{\ddot{z}_{T/S}(t)}{z_{T/S}(t)} \right) v_{k} =0.
\end{aligned}
  \label{MukhanovSasakiCosmicTime}
\end{equation}

When introducing the  new parameter $h_{k}(t)=v_{k}(t)/a(t)$, Eq. (\ref{MukhanovSasakiCosmicTime}) becomes

\begin{equation}
\begin{aligned}
& \ddot{h_{k}}(t)+3H(t)\dot{h_{k}}(t) \\
& + \left[H(t)^{2}+\dfrac{\ddot{a}(t)}{a(t)}+ \mathcal{F}(k_{\varphi})^{2} - H \dfrac{\dot{z}_{T/S}(t)}{z_{T/S}(t)} - \dfrac{\ddot{z}_{T/S}(t)}{z_{T/S}(t)} \right] h_{k}(t) =0.
\end{aligned}
\label{Mukhanov hk Cosmic Time}
\end{equation}

A second parameter $g_{k}(t) = a(t) \dot{h_{k}}(t)$ can also be introduced 
in order to rewrite Eq. (\ref{Mukhanov hk Cosmic Time}) as a set of two first order ODE:

\begin{equation}
\left\{
    \begin{aligned}
& \dot{h_{k}}(t) = \dfrac{1}{a(t)} g_{k}(t)~, \\
& \dot{g_{k}}(t) = - 2 H(t) g_{k}(t) - a(t)h_{k}(t) \times \\
&  \left[ H(t)^{2} + \dfrac{\ddot{a}(t)}{a(t)} + \mathcal{F}(k_{\varphi})^{2} - H(t)\dfrac{\dot{z}_{T/S}(t)}{z_{T/S}} - \dfrac{\ddot{z}_{T/S}(t)}{z_{T/S}} \right] ~. 
\end{aligned}
  \right.
  \label{Set EDO Mukhanov}
\end{equation}

\subsubsection{Equations for tensor modes}
In the case of tensor modes, $z_{T}(t) = a(t)$, and the previous set of ODEs becomes:

\begin{equation}
\left\{
    \begin{aligned}
& \dot{h_{k}}(t) = \dfrac{1}{a(t)} g_{k}(t)~, \\
& \dot{g_{k}}(t) = - 2 H(t) g_{k}(t) - a(t)\mathcal{F}(k_{\varphi})^{2} h_{k}(t) ~. 
\end{aligned}
  \right.
  \label{Set EDO tensor modes RG}
\end{equation}

\subsubsection{Equations for scalar modes}

In the case of scalar modes, $z_{S}(t)=a(t)\dfrac{\dot{\Phi}(t)}{H(t)}$, $\Phi$ being the background variable of the scalar field that fills the universe. Then, the set of Eqs. (\ref{Set EDO Mukhanov}) become:

\begin{equation}
\left\{
    \begin{aligned}
& \dot{h_{k}}(t) = \dfrac{1}{a(t)} g_{k}(t)~, \\
& \dot{g_{k}}(t) = - 2 H(t) g_{k}(t) - a(t) h_{k}(t) \times \\
&  \left[\mathcal{F}(k_{\varphi})^{2} + m^{2} - 2 m^{2} \dfrac{\Phi(t)}{\dot{\Phi}(t)} \dfrac{\dot{H}(t)}{H(t)} \right . \\
& \left . - 2 \left(\dfrac{\dot{H}(t)}{H(t)}\right)^{2} + \dfrac{\ddot{H}(t)}{H(t)} \right], 
\end{aligned}
  \right.
  \label{Set EDo scalar RG}
\end{equation}

As explained in \cite{Schander:2015eja}, the divergence at the bounce, when $H(t)=0$, can be removed by the change of variable $\mathcal{R}(t) = v(t)/z(t)$. The curvature scalar $\mathcal{R}$ is a gauge invariant quantity, related to the Ricci scalar, used to describe the intrinsic curvature.

\subsection{The Mukhanov-Sasaki perturbations equation in the deformed algebra}

The physical motivation for the deformed algebra approach to perturbations in LQC, given in details in \cite{eucl3}, will be recalled later in this article. Here, we just mention how the previous results are modified. \\

The basic logics is the same. The Mukhanov-Sasaki equation in conformal time now reads:
\begin{equation}
v_{k}''(\eta) + \left(\Omega(\eta)k_{c}^{2} - \dfrac{z_{T/S}''(\eta)}{z_{T/S}(\eta)} \right) v_{k}(\eta) =0 ~,
\label{Mukhanov DA}
\end{equation}
with $\Omega(t)=1-2\rho(t)/\rho_c$. The Mukhanov variable $z$ is modified for tensor modes and becomes $z_{T}(\eta)= a(\eta)/\sqrt{\Omega(\eta)}$ whereas $z_{S}$ remains unchanged. The variables $h_{k}$ and $g_{k}$ are redefined as $h_{k} = v_{k}/z$ and $g_{k}=a\dot{h_{k}}/\Omega$, so that the set of first-order differential equations for tensor modes becomes:

\begin{equation}
\left\{
    \begin{aligned}
& \dot{h_{k}}(t) = \dfrac{\Omega(t)}{a(t)} g_{k}(t)~, \\
& \dot{g_{k}}(t) = - 2 H(t) g_{k}(t) - a(t)\mathcal{F}(k_{\varphi})^{2}h_{k}(t) ~. 
\end{aligned}
  \right.
  \label{Set EDO Mukhanov DA}
\end{equation}

For scalar modes there is no need to redefine $h_{k}$ and $g_{k}$. The new set of equations is:

\begin{widetext}
\begin{equation}
\left\{
    \begin{aligned}
& \dot{h_{k}}(t) = \dfrac{1}{a(t)} g_{k}(t)~, \\
& \dot{g_{k}}(t) = - 2 H(t) g_{k}(t) - a(t) h_{k}(t) \times \\
&  \left[\Omega(t)\mathcal{F}(k_{\varphi})^{2} + m^{2} + \kappa m^{2} (1- 2 \rho(t)/\rho_{c}) \dfrac{\Phi(t)\dot{\Phi}(t)}{H(t)} - 2 \left(\dfrac{\dot{H}(t)}{H(t)}\right)^{2} + \dfrac{\ddot{H}(t)}{H(t)} \right], 
\end{aligned}
  \right.
\end{equation}
\end{widetext}
where the difference with Eq.(\ref{Set EDo scalar RG}) also comes from the appearance of the $\Omega$ factor, this time in the second equation. The expression for the power spectrum $\mathcal{P}_{S}(k_{c})$ remains unchanged.

\subsection{Initial conditions for perturbations}

In this study, and following the logics of causality, initial conditions are set in the contracting branch, where the MDR effects vanish, so that $a(\eta)\mathcal{F}(k_{\varphi}) \simeq a(\eta)k_{\varphi} = k_{c}$.

Moreover, the tensor effective potential term $z_{T}''/z_{T}$ tends to zero in the contracting phase both in the standard view and in the deformed algebra approach. It is therefore possible to find, for every comoving wavenumber $k_{c}$, a time $\eta_{i}$ in the contracting phase such that $k_{c} \gg z_{T}''(\eta_{i})/z_{T}(\eta_{i})$. As long as this assumption is valid, solutions to Eq. (\ref{MukhanovSasaki Temporal modes physical k}) can be written:

\begin{equation}
v_{k}(\eta) = c_{1} e^{ik_{c}\eta} + c_{2} e^{-ik_{c}\eta} ~.
\end{equation}

The two constants $c_{1}$ and $c_{2}$ are constrained by the Wronskian condition:

\begin{equation}
v_{k} \dfrac{dv_{k}^{\star}}{d\eta} - v_{k}^{\star} \dfrac{dv_{k}}{d\eta} = i ~,
\end{equation}
which comes from relations on the commutators
$\left[\hat{a}_{\vec{k}},\hat{a}^{\dagger}_{\vec{q}}\right] = \delta^{(3)}(\vec{k}-\vec{q})$ introduced for the temporal mode functions operator in the Heinseberg picture: $\hat{v}_{\vec{k}}(\eta) = v_{k}(\eta) \hat{a}_{\vec{k}} + v_{k}^{\star}(\eta) \hat{a}^{\dagger}_{-\vec{k}}$ .
This condition implies:

\begin{equation}
\vert c_{2} \vert ^{2} - \vert c_{1} \vert ^{2} = \dfrac{1}{2k_{c}}.
\end{equation}

Choosing $c_{1} = 0 $ and $c_{2} = 1/\sqrt{2k_{c}}$ in order to describe a wave propagating in the positive time direction, the mode functions  read:

\begin{equation}
v_{k}(\eta) = \dfrac{1}{\sqrt{2k_{c}}} e^{- i k_{c} \eta } ~~~.
\end{equation}

This normalization, called \textit{Minkowski Vacuum}, can be transposed to the $h_{k}$ and $g_{k}$ coefficients:

\begin{equation}
\left\{
    \begin{aligned}
& h_{k}(t_{i}) = \dfrac{1}{a(t_{i})} \dfrac{1}{\sqrt{2k_{c}}}, \\
& g_{k}(t_{i}) = \dfrac{- i}{a(t_{i})} \dfrac{\sqrt{k_{c}}}{\sqrt{2}} - H(t_{i}) \dfrac{1}{\sqrt{2k_{c}}}. 
\end{aligned}
  \right.
\end{equation}

The case of scalar modes is slightly more delicates due to the $z_{S}''/z_{S}$ scalar potential behaviour in the contracting branch. In this regime $z_{S}''/z_{S} \simeq - m^{2} a(t)^{2}$. The potential is proportional to the scale factor squared and cannot be neglected anymore. This issue has already been addressed in \citep{Schander:2015eja} and initial conditions were set using a WKB approximation and fixing a precise time at which this approximation is valid. We have applied the same strategy in this study. It is probably not the final word on this issue and the infrared behavior of the scalar spectra, directly sensitive to the initial normalization, must not be considered too seriously. However, since MDR effects appear in the UV regime, this is not a major concern for this study.\\

\section{Unruh-like Modified Dispersion Relation}

\subsection{Physical justification}

%This MDR has initially been introduced by Unruh !!!ref!!! to determine the high frequency regime of black hole's thermal emission. The emitted radiation depends on this regime, in which quantum effects of gravitation must be taken into accounts. Since there was (there still is?) no satisfying theory of quantum gravity, quantum gravitational effects had been implemented wih the use of a MDR. This was done using the analogy between sonic waves propagating in a hypersonic fluid and scalar perturbations propagating in a black hole spacetime, which permitted Unruh to use Dumb holes, the sonic analogous to black holes, for his study. In the case of dumb holes, for wavelengths shorter than the intermolecular space, the continuum fluid approximation is no longer valid  and propagation equations for sonic waves deviate from the usual case.\\

We will first consider the MDR introduced by Unruh in \cite{Unruh:1994je}. The idea was to investigate the sensitivity of the naive calculation of the black hole evaporation spectra to possible modified dispersion relation at high frequencies. 

\begin{figure}[!h]
\begin{center}
\includegraphics[scale=0.55]{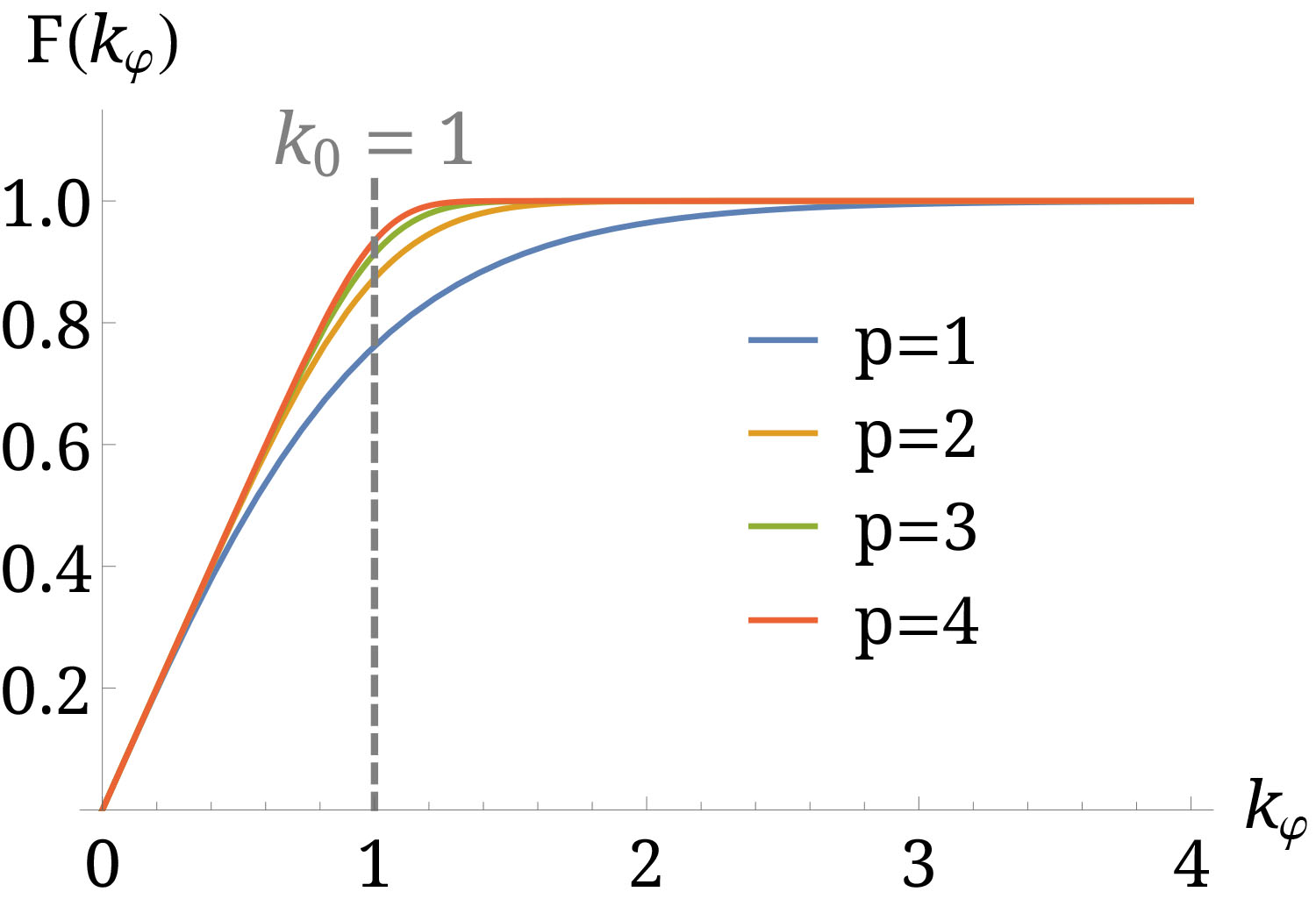}
\caption{Unruh-like dispersion relation for different values of $p$ and $k_{0}=1$.} 
\label{Plot Unruh MDR}
\end{center}
\end{figure}

This MDR writes:

\begin{equation}
\mathcal{F}(k_{\varphi}) = k_{0} \tanh\left[\left(\dfrac{k_{\varphi}}{k_{0}}\right)^{p}\right]^{\dfrac{1}{p}} = k_{0} \tanh\left[\left(\dfrac{k_{c}}{a(t)k_{0}}\right)^{p}\right]^{\dfrac{1}{p}},
\end{equation}

where $k_{0}$ is a physical wavenumber, expected to be of the order of the Planck value, which determines the transition scale at which $\mathcal{F}(k_{\varphi})$ switches from a linear behaviour in $k_{\varphi}$ to a constant. The $p$ parameter determines the sharpness of the transition, as it can be seen on Fig.\ref{Plot Unruh MDR}.

Clearly, our problematic is quite different from the one initially considered by Unruh. This MDR might however phenomenologically capture some quantum gravity effects. When the ``effective" physical wavenumber becomes trans-planckian, the energy does not grow anymore. Intuitively this can be thought of as a kind of saturation taking place at the Planck scale. It is important to stress once again that this might happen when the Universe {\it density} is far from planckian. In the contracting branch, for example, the wavelength of a given mode can reach the Planck size when the background density is still very small and the dynamics fully described by the classical (contracting) Friedmann equation. 

\subsection{Primordial power spectra}

\subsubsection{Deformed algebra approach}

\begin{figure}[!h]
\begin{center}
\includegraphics[scale=1.10]{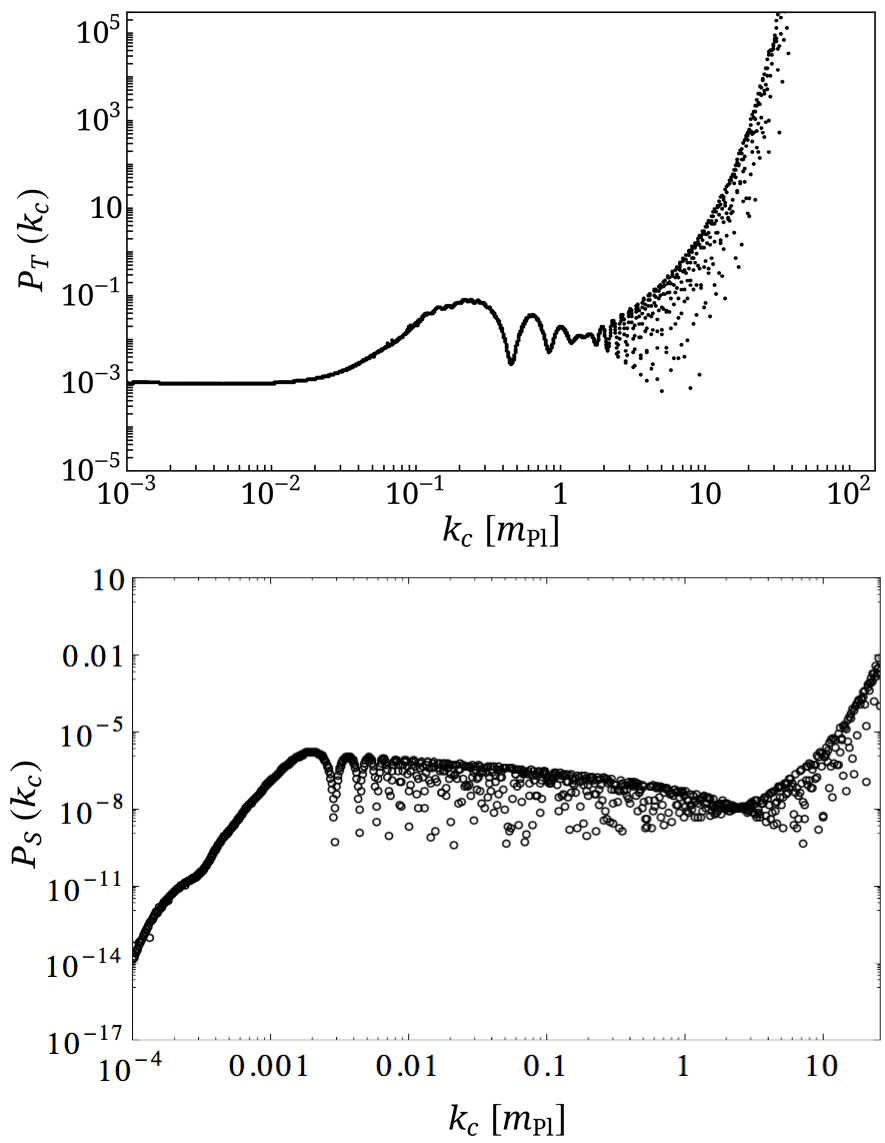}
\caption{Upper: Primordial tensor power spectrum. Lower: Primordial scalar power spectrum. Both spectra are obtained in the Deformed Algebra approach and with the usual dispersion relation $\mathcal{F}(k_{\varphi}) = k_{\varphi}$.} 
\label{DA-NoMDR}
\end{center}
\end{figure}

On Fig. \ref{DA-NoMDR} we recall the results previously obtained \cite{lcbg,Bolliet:2015bka,Schander:2015eja} on the tensor and scalar power spectra in the deformed algebra approach. The spectra are respectively defined by:

\begin{equation}
\mathcal{P}_{T}(k_{c}) = \dfrac{4 \kappa k^{3}}{\pi^{2}} \left|\dfrac{v_{k}(\eta_{e})}{z_{T}(\eta_{e})} \right|^{2},
\end{equation}

and

\begin{equation}
\mathcal{P}_{S}(k_{c}) = \dfrac{k^{3}}{2 \pi^{2}} \left|\dfrac{v_{k}(\eta_{e})}{z_{S}(\eta_{e})} \right|^{2},
\end{equation}

where $\eta_{e}$ stands for the conformal time at the end of the inflationary period.
For all scalar spectra presented in this article the scalar field mass is fixed at $m=1.2\times 10^{-6} m_{Pl}$, as favored by Planck data \citep{Planck2015}. However, since tensor perturbations are only metric fluctuations, the scalar field mass do not appear explicitly in the Mukhanov-Sasaki equation and it has been shown in \citep{Bolliet:2015bka} that a modification of this mass does not modify the spectrum shape. In order to decrease the simulation running time, we have therefore set $m=1.2\times 10^{-2} m_{Pl}$ for all the tensor spectra.\\ 

The deformed algebra approach puts a specific emphasis on gauge issues that are critical in gravity (it is not a priori clear whether the first class nature of the system of constraints can persist when leading order quantum corrections are included, see \cite{Ashtekar:2015dja}). In the constraints, the gravitational connection is basically replaced by its holonomy. The quantum-corrected constraints are calculated for the perturbations up to the desired order and the Poisson brackets are evaluated. Anomalies are cancelled by suitable counter-terms, which are required to vanish in the classical limit. The resulting theory is not only anomaly free  but is also uniquely defined when matter is included \cite{tom1,tom2,eucl2}. Although the calculations involve intricate expressions, the resulting final algebra is simple and depends on a unique structure function which encodes all the modifications: $\Omega = 1 - 2\rho/\rho_c$. The resulting spectra, at least when perturbations are propagated through the bounce (other possibilities are studied in \cite{Bojowald:2015gra,Mielczarek:2014kea,Barrau:2016sqp}) exhibit three different regimes: either a scale-invariant behavior or a growth $\propto k^{3}$ in the IR\footnote{The IR behavior of the scalar mode spectrum is highly dependent on the way to set initial conditions and remains an open issue.}, depending on the nature of the perturbations, oscillations in the intermediate part and an exponential divergence in the UV. As, such, those spectra are excluded by data \cite{Bolliet:2015raa}.

\begin{figure}[!h]
\begin{center}
\includegraphics[scale=0.40]{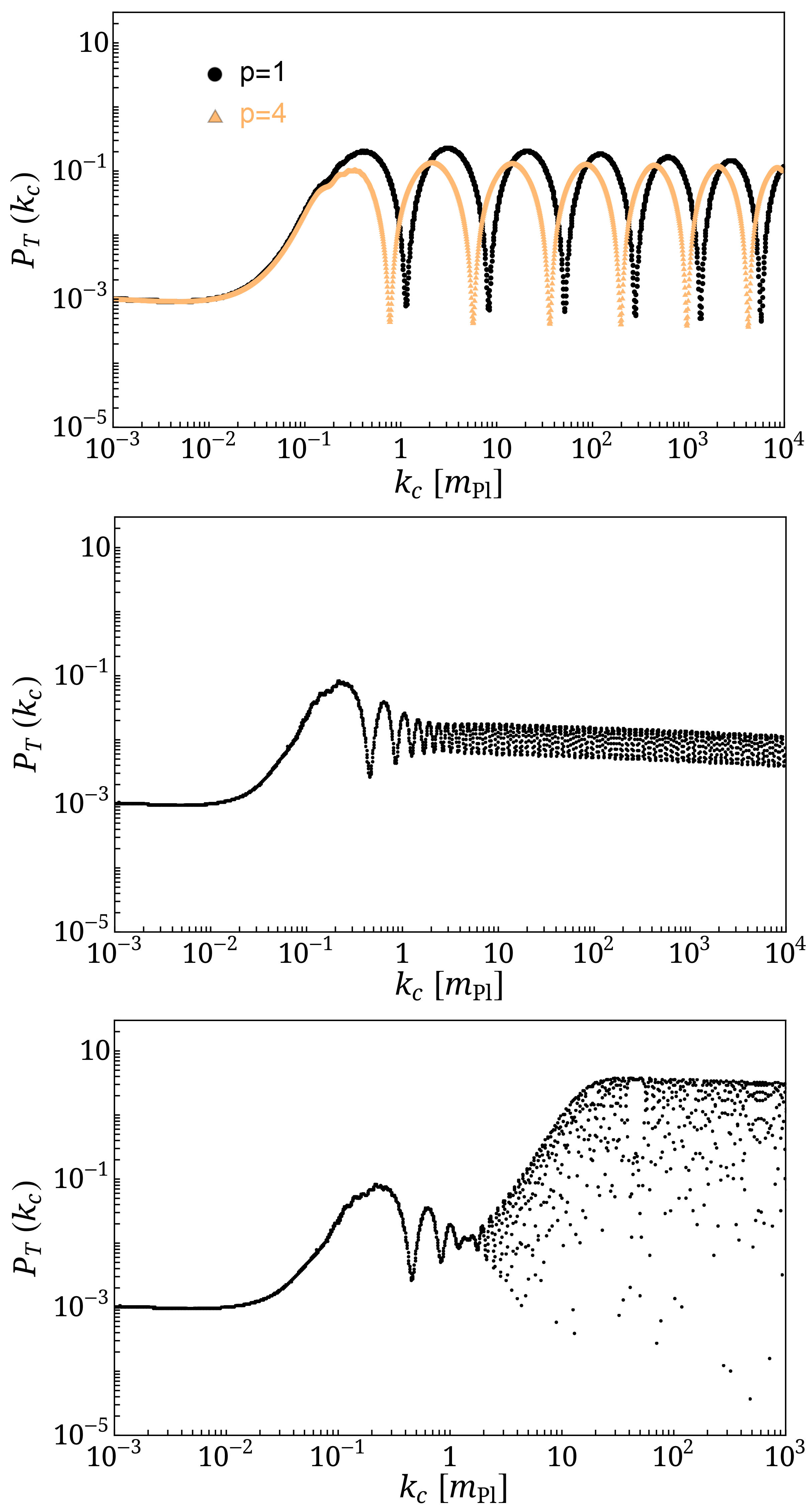}
\caption{Primordial tensor power spectra obtained in the Deformed Algebra approach with the Unruh-like MDR $\mathcal{F}(k_{\varphi}) = k_{0} \tanh\left[\left(k_{c}/(a(t)k_{0})\right)^{p}\right]^{1/p}$, for different $k_{0}$ and $P$ values. Upper: $k_{0}=0.1$, $p=1$ and $p=4$. Middle: $k_{0}=1$, $p=1$. Lower: $k_{0} = 10$, $p=1$.} 
\label{DA-Tensor-Unruh}
\end{center}
\end{figure}

\begin{figure}[!h]
\begin{center}
\includegraphics[scale=1.10]{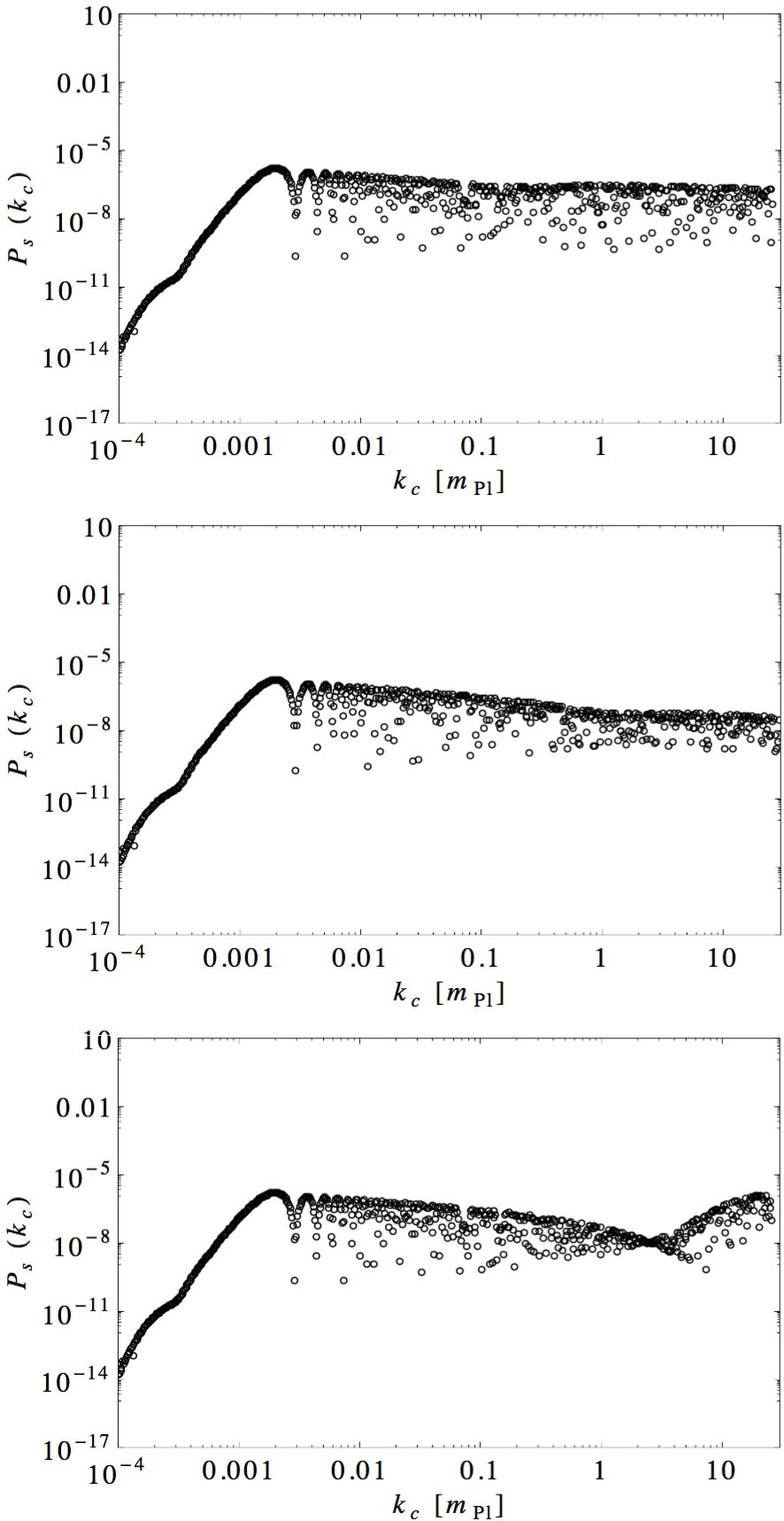}
\caption{Primordial scalar power spectra obtained in the Deformed Algebra approach with the Unruh-like MDR $\mathcal{F}(k_{\varphi}) = k_{0} \tanh\left[\left(k_{c}/(a(t)k_{0})\right)^{p}\right]^{1/p}$, for $p=1$ and different $k_{0}$ values. Upper: $k_{0}=0.1$. Middle: $k_{0}=1$. Lower: $k_{0} = 10$.} 
\label{DA-Scalar-Unruh}
\end{center}
\end{figure}

The shape of the spectra after implementing the Unruh MDR, as shown on Fig. \ref{DA-Tensor-Unruh} for tensor modes and Fig. \ref{DA-Scalar-Unruh} for scalar modes, can be quite easily understood. For simplicity we skip the explicit time-dependence of the modes. Around the bounce, $\Omega$ is negative and the equation of motion can be written as

\begin{equation}
v''- ( |\Omega| a^2 \mathcal{F}^2(k_{\varphi}) +z''/z) v = 0.
\label{expl1}
\end{equation}
Assuming that close to the bounce, $z''/z$ varies relatively slowly and can be approximated by a time-independent constant, as a crude approximation which should just be considered as a way to get an intuition of the main behavior, one obtains (keeping only the growing solution):

\begin{equation}
v \approx A e^{|\Omega| a^2 \mathcal{F}^2(k_{\varphi}) +z''/z } \approx B e^{ |\Omega| a^2 \mathcal{F}^2(k_{\varphi}) }.
\label{expl2}
\end{equation}

When no MRD is implemented, $a^2\mathcal{F}^2(k_{\varphi})=k_{c}^2$ and one can easily observe the exponential behaviour. With the Unruh's MDR, $\mathcal{F}(k_{\varphi})\approx k_{0}$ when $k> k_{0}$ and one simply expect a constant behaviour. 

The upper panel of Fig. \ref{DA-Tensor-Unruh} also shows that the precise value of $p$ plays, as expected, no significant role. We therefore fix it to $p=1$ in the other plots.\\

Importantly, this study shows that when an Unruh-like MRD is used the pathology exhibited by the deformed algebra spectrum is fully cured as long as the ``transition" wavenumber $k_{0}$ is not taken at un-naturally high value. Trans-planckian effects can therefore drastically modify the UV part of the primordial power spectra.

\subsubsection{Dressed metric approach}

\begin{figure}[!h]
\begin{center}
\includegraphics[scale=1.10]{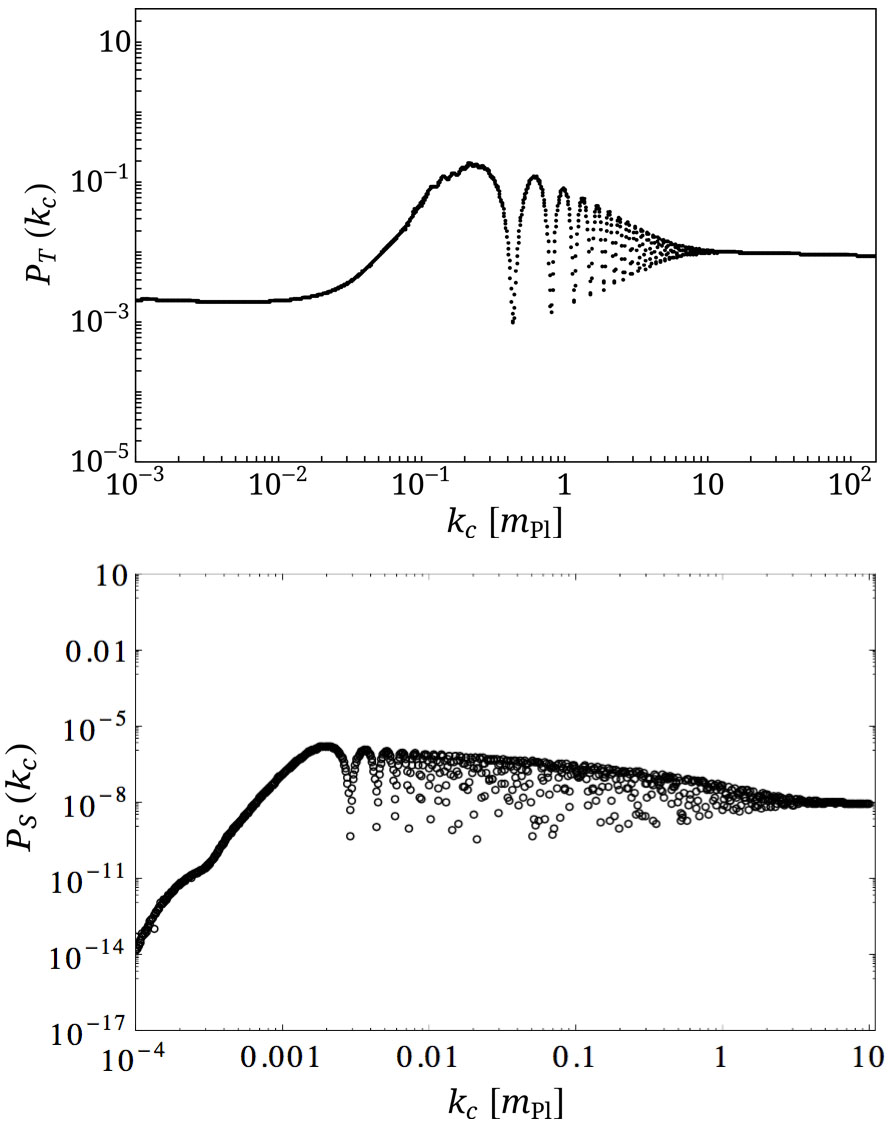}
\caption{Upper: Primordial tensor power spectrum. Lower: Primordial scalar power spectrum. Both spectra are obtained in the Dressed Metric approach and with the standard dispersion relation $\mathcal{F}(k_{\varphi}) = k_{\varphi}$.} 
\label{DM-NoMDR}
\end{center}
\end{figure}

The dressed metric approach \cite{Agullo1,Agullo2,Agullo3}  deals with quantum fields on a quantum background. 
It relies on a minisuperspace strategy where the homogeneous and inhomogeneous degrees of freedom are both quantized. The former quantization follows the loop approach whereas the latter is performed thanks to a Fock-like procedure. The physical perturbations are given by the Mukhanov-Sasaki variables derived from the linearised classical constraints. The full Hilbert space is the tensor product of the Hilbert space for the background with the Hilbert space of the perturbed degrees of freedom. The Schr\"odinger equation for the perturbations was shown to be formally identical to the Schr\"odinger equation for the quantized perturbations evolving on a classical background and feeling a dressed metric that takes into account the quantum nature of this background. The full consistency is however not yet established in the sense that it is not clear that the constraints still constitute a ``first class" system. If this important symmetry of the classical theory is lost one may obtain meaningless (gauge-dependent) results. Gauge-fixing before quantization is often harmless in physics but the case of gravity is peculiar as dynamics is part of the gauge system) \cite{eucl3}.\\

\begin{figure}[!h]
\begin{center}
\includegraphics[scale=0.40]{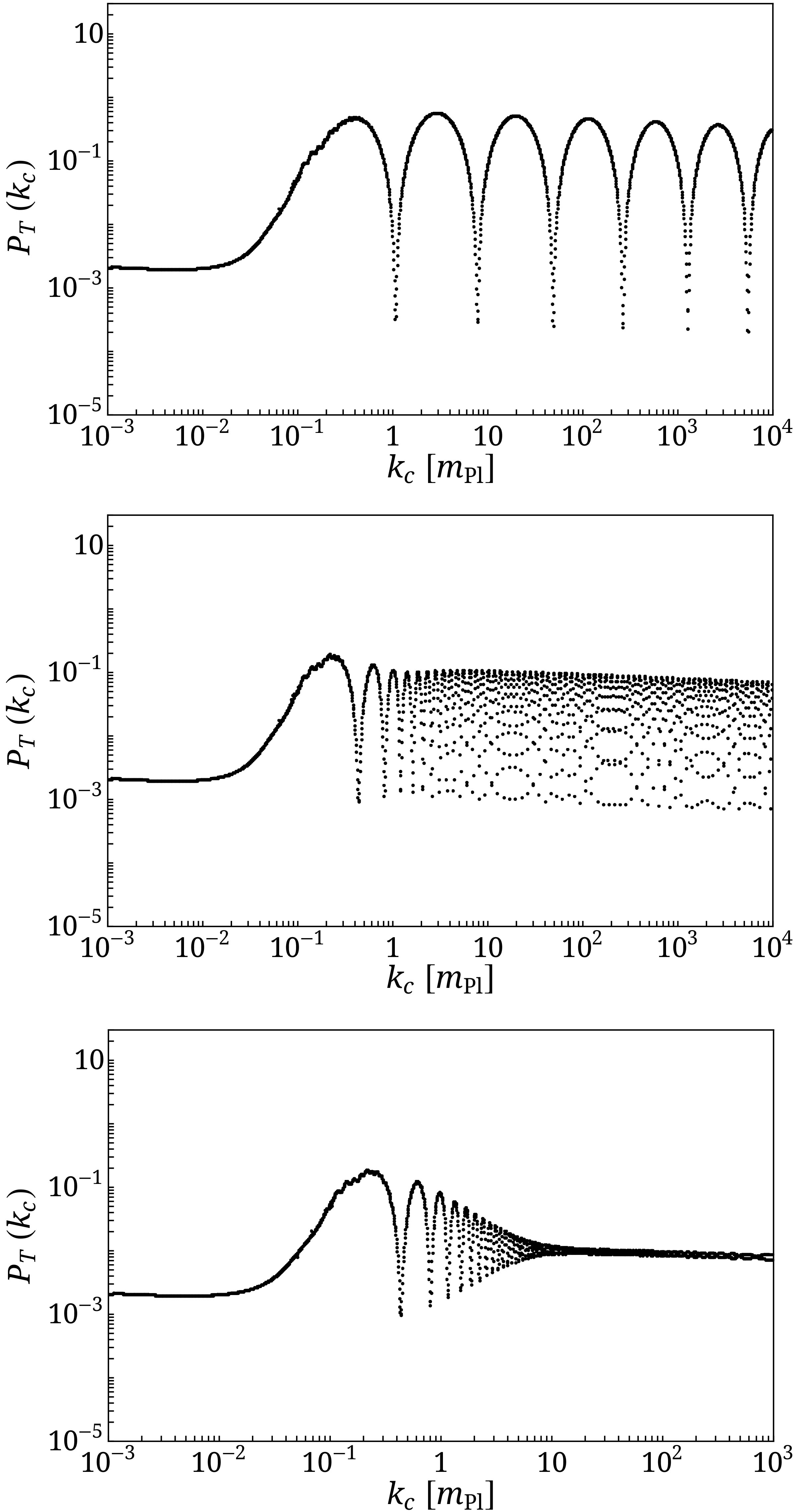}
\caption{Primordial tensor power spectra obtained in the Dressed Metric approach with the Unruh-like MDR $\mathcal{F}(k_{\varphi}) = k_{0} \tanh\left[\left(k_{c}/(a(t)k_{0})\right)^{p}\right]^{1/p}$, for $p=1$ and different $k_{0}$ values. Upper: $k_{0}=0.1$. Middle: $k_{0}=1$. Lower: $k_{0} = 10$.} 
\label{DM-Tensor-Unruh}
\end{center}
\end{figure}

\begin{figure}[!h]
\begin{center}
\includegraphics[scale=1.10]{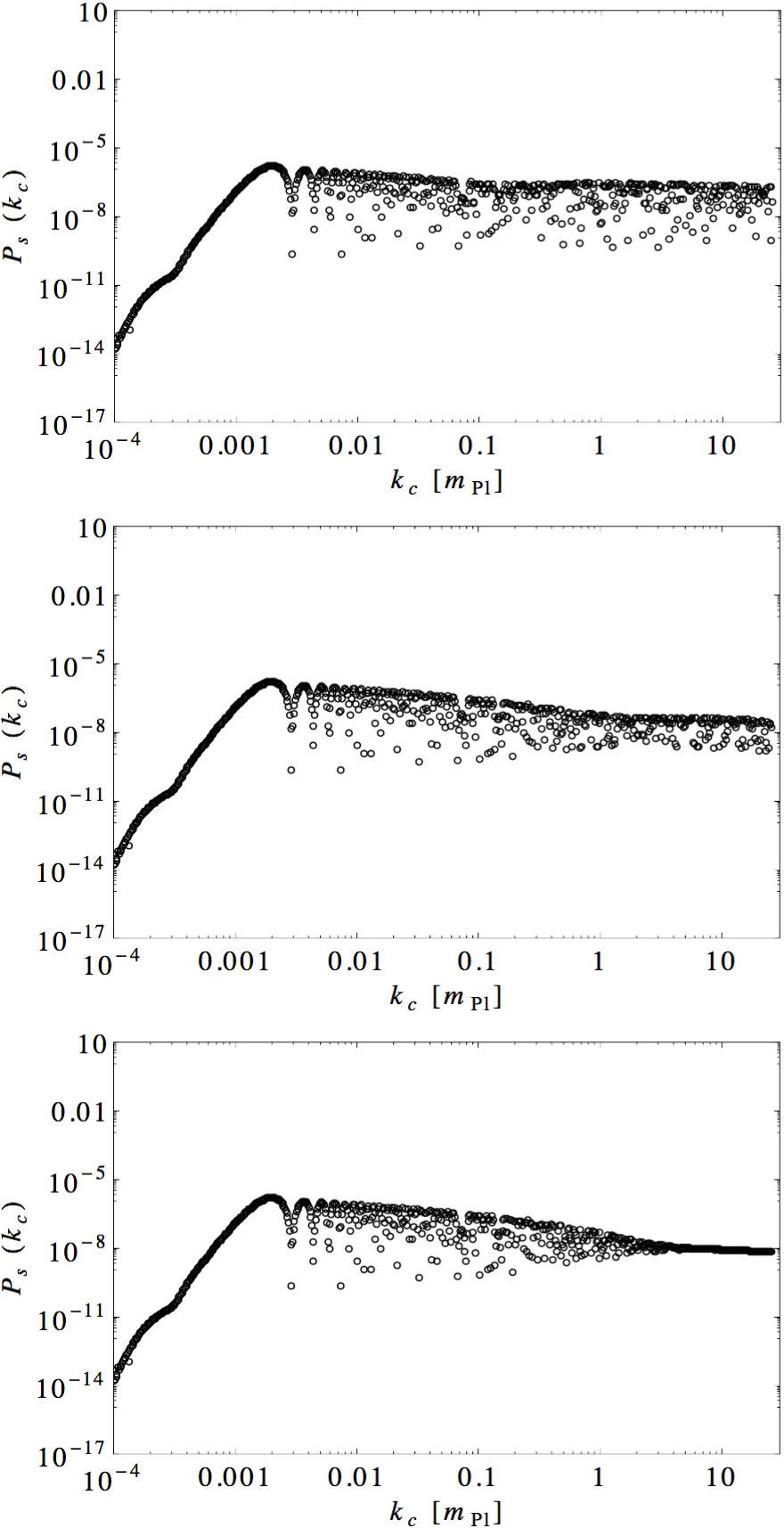}
\caption{Primordial scalar power spectra obtained in the Dressed Metric approach with the Unruh-like MDR $\mathcal{F}(k_{\varphi}) = k_{0} \tanh\left[\left(k_{c}/(a(t)k_{0})\right)^{p}\right]^{1/p}$, for $p=1$ and different $k_{0}$ values. Upper: $k_{0}=0.1$. Middle: $k_{0}=1$. Lower: $k_{0} = 10$.} 
\label{DM-Scalar-Unruh}
\end{center}
\end{figure}

Interestingly, for sharply peaked states, this approach basically leads to the same equation of motion for the perturbations than the usual GR one. The results given here (and in the ``dressed metric" paragraph of the next section) can therefore either be considered as showing the MDR effects on the dressed metric approach itself or as the MDR effects on standard perturbations moving in a LQC-like background.\\

As seen on Figs. \ref{DM-Tensor-Unruh} and \ref{DM-Scalar-Unruh}, the implementation of this MDR does not change the dressed metric nearly scale-invariant UV power spectrum obtained without MDR, presented on Fig.\ref{DM-NoMDR}. \\

There are however few quantitative features induced by the MDR that are worth to be mentioned. For the tensor modes first, the frequency of the oscillations can become rather large for small values of $k_0$, as clearly seen on the upper panel of Fig. \ref{DM-Tensor-Unruh} for $k_0=0.1$. In full generality, any feature in the primordial power spectrum can be transferred into features in the angular power spectra, $C_\ell$, of the cosmic microwave background anisotropies. Without MDR however, the oscillations in the $\mathcal{P}_T$ are washed out because the transfer functions (acting as a kernel translating primordial inhomogeneities into CMB anisotropies) are wide enough in the $k$-space, as compared to the oscillations, to smooth them. With an Unruh-like MDR and assuming a small-enough value of $k_0$, this may not be the case anymore for the oscillations becomes wide enough (for example with $k_0=0.1$, the frequency of oscillations is first roughly constant per logarithmic band with a rough value of 1, {\it i.e.} there is one oscillation per decade in $k$).

For the scalar modes then, the prediction without MDR for wavenumbers greater than few $m_{Pl}$ is a slightly red-tilted spectrum because the amplification of these modes is dominated by the inflationary phase following the bounce. As can be seen on Fig. \ref{DM-Scalar-Unruh}, accouting for the MDR slightly reduces the steepness of this tilt for $k>k_0$. This can be understood as follows: with the Unruh-like MDR, the primrodial power spectrum for $k>k_0$ is roughly such that $\mathcal{P}_S(k>k_0)\sim\mathcal{P}_S(k_0)$. 

At this stage, it is however unclear if the such quantitative differences are large enough to be tested.

\section{Exponentially suppressed Dispersion Relation}

\subsection{Physical justification}

\begin{figure}[!h]
\begin{center}
\includegraphics[scale=0.55]{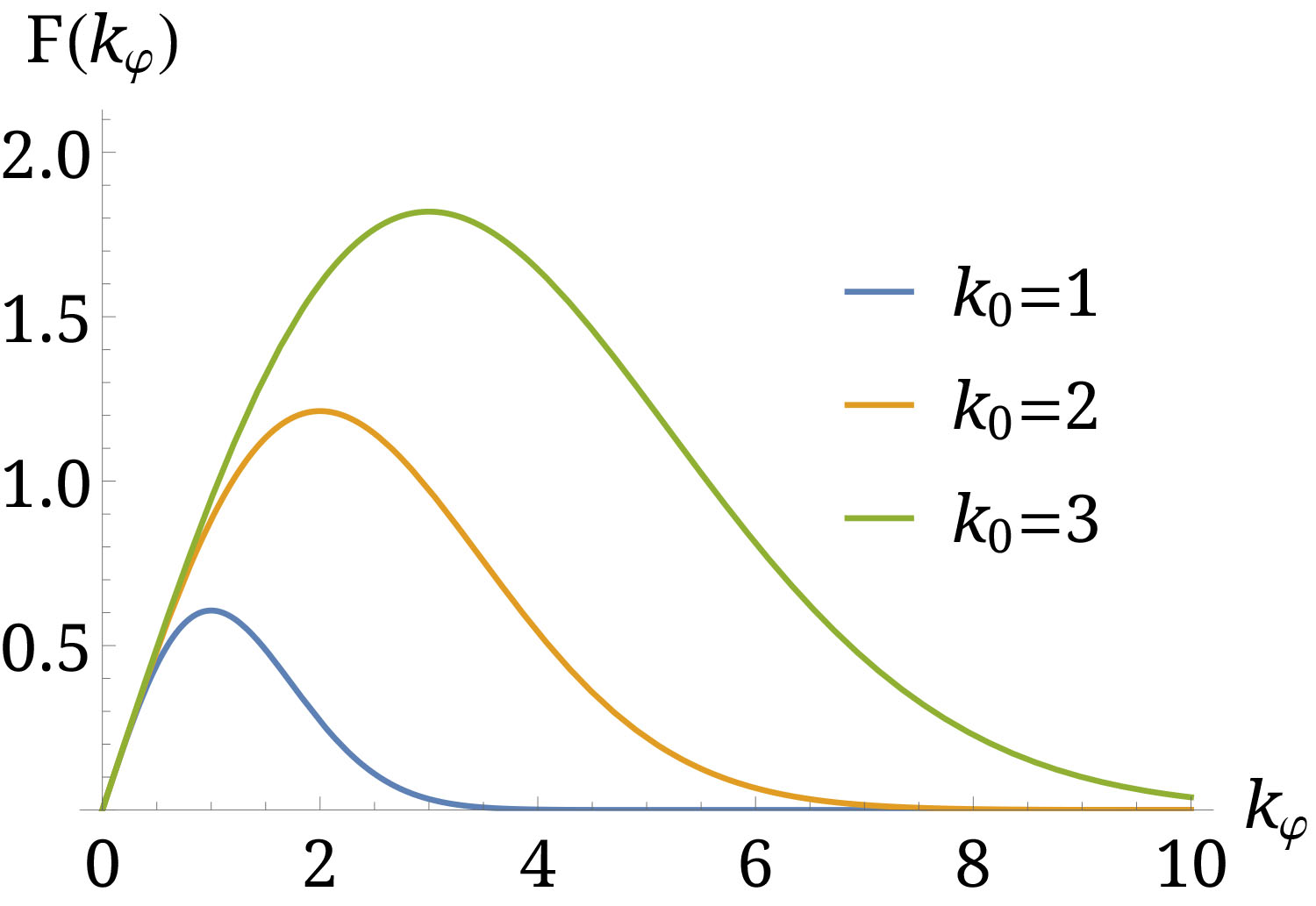}
\caption{Exponentially suppressed dispersion relation for different values of $k_{0}$.} 
\label{Plot MDR Expo}
\end{center}
\end{figure}

We consider here another MDR given by 

\begin{equation}
\mathcal{F}(k_{\varphi})^{2} = k_{\varphi}^{2}e^{-\frac{k_{\varphi}^{2}}{k_{0}^{2}}}.
\label{MDR expo}
\end{equation}
It was explicitly used in \cite{Lemoine:2001ar}, inspired by a slightly different proposal made in \cite{MersiniHoughton:2001su}. This relation is, as it should be, linear in the IR. It reaches a maximum around $k_0$ and then decreases to zero. The physical interpretation of this MDR is different from Unruh's one. Here, a vanishing energy is associated with modes characterized by an effective trans-planckian wavenumber, so as to account for their ``effective" disappearance. The shape of $\mathcal{F}(k_{\varphi})$ is shown on Fig. \ref{Plot MDR Expo}. This kind of MDRs also appears in effective string theory as a consequence of the T-duality symmetry (equivalence between strings compactified around a circle of radius $R$ and $1/R$). In this case the string length $l_s$ directly enters the formula and some cosmological consequences were considered in \cite{Chouha:2005yn}.

\subsubsection{Deformed algebra approach}

\begin{figure}[!h]
\begin{center}
\includegraphics[scale=0.40]{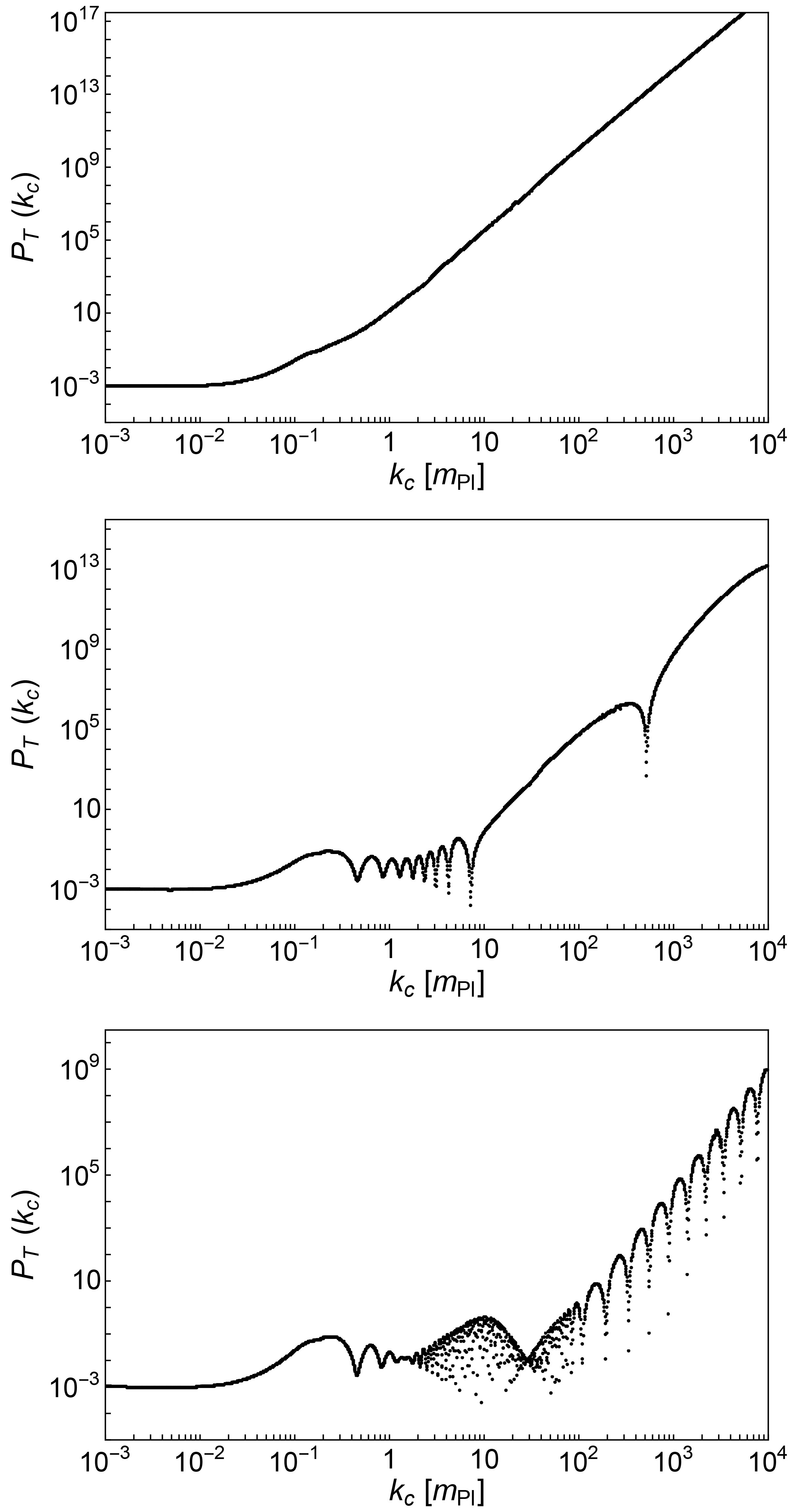}
\caption{Primordial tensor power spectra obtained in the Deformed Algebra approach with the exponentially-suppressed MDR $\mathcal{F}(k_{\varphi})^{2} = k_{\varphi}^{2}e^{-k_{\varphi}^{2}/k_{0}^{2}}$, for different $k_{0}$ values. Upper: $k_{0}=0.1$. Middle: $k_{0}=1$. Lower: $k_{0} = 10$.} 
\label{DA-Expo}
\end{center}
\end{figure}

As it can be seen on Fig.\ref{DA-Expo}, this MDR leads to a very blue UV spectrum. Intuitively one can interpret this behavior as a kind of ``remapping" of the spectrum: the function $\mathcal{F}(k_{\varphi})$ replaces $k_{\varphi}$ in the equations of motion. But $\mathcal{F}(k_{\varphi})$ is now decreasing with $k_{\varphi}$ in the trans-planckian regime. It means that the shape of spectrum when going to higher values of $k_{\varphi}$, beyond $k_{0}$, is expected to mimic the (possibly stretched) behavior of the spectrum when going to smaller values of $k_{\varphi}$ below $k_{0}$. The observed rise for wavenumbers greater than $k_0$ is therefore connected with the red envelop of the oscillations of the spectrum for wavenumbers smaller than $k_0$.\\

In this section (and in the following one), we only show the tensor spectrum because, as previously mentioned, the IR behaviour of the scalar spectrum (which is quite controversial as it heavily depends on the way to set initial conditions) would be reflected in the UV through the MDR. There is therefore no point studying it in details here.

\subsubsection{Dressed metric approach}

\begin{figure}[!h]
\begin{center}
\includegraphics[scale=0.40]{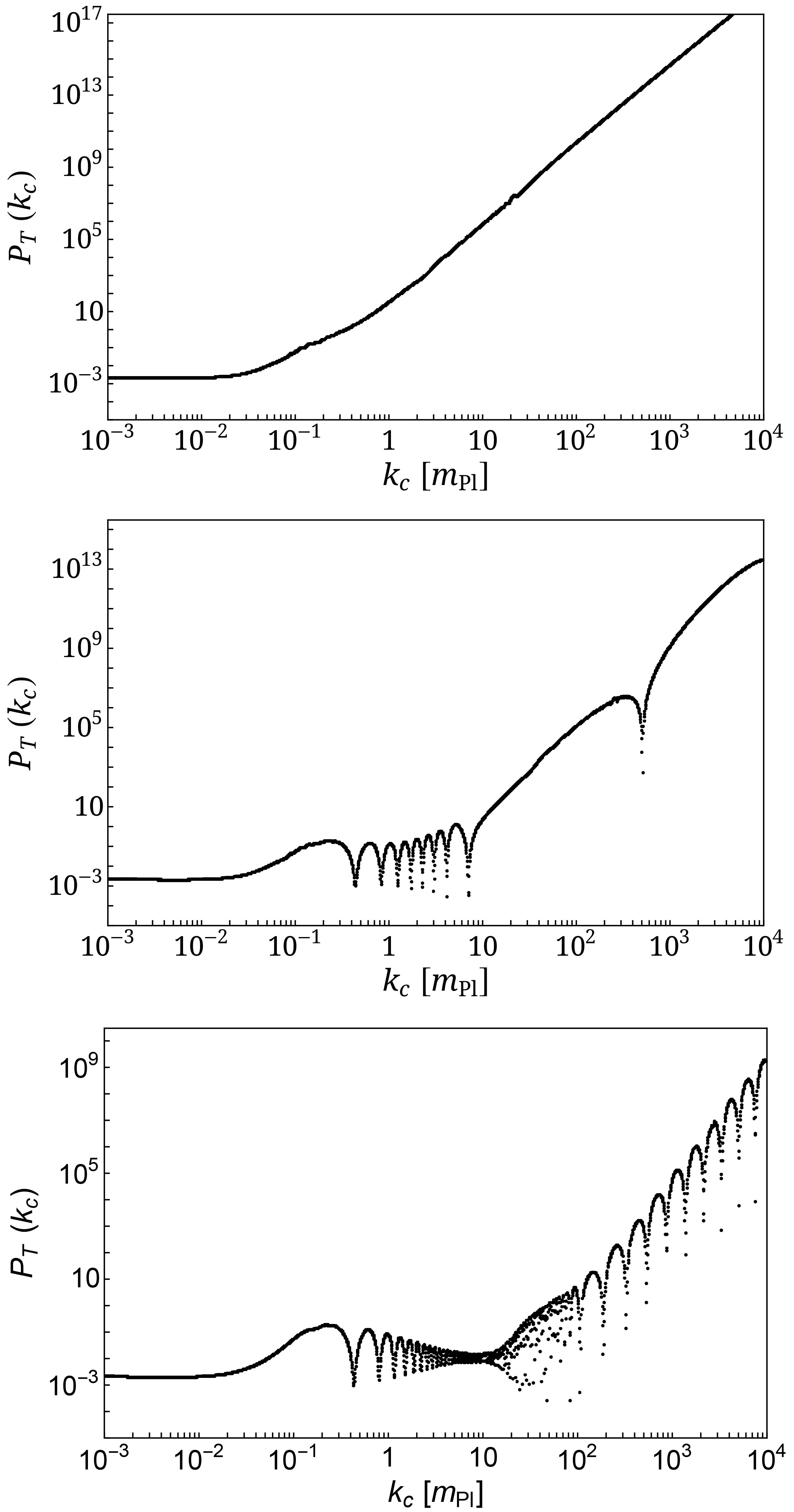}
\caption{Primordial tensor power spectra obtained in the Dressed Metric approach with the exponentially-suppressed MDR $\mathcal{F}(k_{\varphi})^{2} = k_{\varphi}^{2}e^{-k_{\varphi}^{2}/k_{0}^{2}}$, for different $k_{0}$ values. Upper: $k_{0}=0.1$. Middle: $k_{0}=1$. Lower: $k_{0} = 10$.} 
\label{DM-Expo}
\end{center}
\end{figure}

Fig.\ref{DM-Expo} shows, for the reason given in the previous section, that the spectrum does now exhibit a deeply blue behaviour at small scales, as for the deformed algebra spectrum. This is interesting as it was believed that quantum gravity effects were, in this framework, only affecting the IR modes. The reason for that was that those modes are, at the bounce time, larger than the curvature radius of the Universe and are therefore ``feeling" the curvature. They are thus excited and not in the Bunch-Davies vacuum. This quantum-gravity effect was only due to the background dynamics and therefore to the density. In the case of the MDR here implemented, the physical size of the mode also plays a role and the UV part of the spectrum is affected.

\section{Conclusion}

In the framework of loop quantum cosmology, the background behavior when the density becomes planckian is quite well known and understood. The effects of the resulting bouncing dynamics on the spectrum of primordial perturbations have also been intensively studied. In this article, we have focused on another issue: the consequences of quantum gravity effects on the {\it length} of the considered mode, independently of the {\it density} of the background. 

As a first step in this direction, we have considered two different modified dispersion relations to take into account the existence of a minimum scale in loop quantum gravity: the Unruh's one and the exponentially suppressed one. The first lesson of this study is that the trans-planckian behaviour of the MDR does matter and changes the spectrum. In particular:
\begin{itemize}
\item the pathological behavior of the ``naive" ({\it i.e.} with a quadratic potential and no backreaction) deformed algebra spectrum can be cured by an Unruh-like MDR.
\item the correct behavior of the dressed metric spectrum can become pathological if an exponentially decreasing MDR is implemented.
\end{itemize}
This makes the situation particularly interesting in the sense that the anomaly-free approach is now shown, when trans-planckian effects are taken into account, to be possibly in agreement with data. But this does not mean that the spectra are insensitive to the details of quantum gravity effects as some MDRs are clearly incompatible with measurements. Those effects cannot be ignored as, for a typical duration of inflation, the modes observed in the current Universe were deeply trans-planckian at the bounce time.\\

In the black hole sector \cite{AmelinoCamelia:2005ik,Ling:2005bq,Tao:2017mpe}, it is now known that MDRs have a very small impact on the characteristics of the Hawking radiation. This means that the predictions are quite insensitive to trans-planckian physics. This is also true in the cosmological sector \cite{Starobinsky:2001kn} when dealing with the usual inflationary paradigm. However, we have shown here that this is not the case anymore when dealing with a quantum-gravity induced bouncing scenario (as already suggested in \cite{Brandenberger:2002ty}). This obviously means that the predictions are less robust but this also opens an interesting window on trans-plackian physics.

As a next step in this direction, it would be important to use less {\it ad hoc} hypotheses for the shape of the MDRs but, instead, to derive them from the full theory. Although the task is quite hard if taken fully rigorously, some toy-models inspired by solid-state physics approaches can certainly be built.

\section*{Acknowledgments}

K.M is supported by a grant from the CFM foundation.

\bibliography{refs}
\end{document}